\documentclass[aps,prl,superscriptaddress,twoside,twocolumn,nofootinbib,showpacs,floatfix]{revtex4-1}
\usepackage{amsmath,amssymb}
\usepackage{graphicx}
\usepackage{braket,bm}
\usepackage{subfigure}
\allowdisplaybreaks
\begin{document}


\title{Is $d^*$ a candidate of hexaquark-dominated exotic state?}

\author{F. Huang}
\affiliation{School of Physical Sciences, University of Chinese Academy of Sciences, Beijing 101408, China}

\author{Z.Y. Zhang}
\affiliation{Institute of High Energy Physics, Chinese Academy of Sciences, Beijing 100049, China}
\affiliation{Theoretical Physics Center for Science Facilities, CAS, Beijing 100049, China}

\author{P.N. Shen}
\affiliation{Institute of High Energy Physics, Chinese Academy of Sciences, Beijing 100049, China}
\affiliation{Theoretical Physics Center for Science Facilities, CAS, Beijing 100049, China}
\affiliation{College of Physics and Technology, Guangxi Normal University, Guilin 541004, China}

\author{W.L. Wang}
\affiliation{School of Physics and Nuclear Energy Engineering, Beihang University, Beijing 100191, China}

\date{\today --- \jobname}


\begin{abstract}
We confirm our previous prediction of a $d^*$ state with $I(J^P)=0(3^+)$ [Phys. Rev. C {\bf 60}, 045203 (1999)] and report for the first time based on a microscopic calculation that $d^*$ has about 2/3 hidden color (CC) configurations and thus is a hexaquark-dominated exotic state. By performing a more elaborate dynamical coupled-channels investigation of the $\Delta\Delta$-CC system within the framework of resonating group method (RGM) in a chiral quark model, we found that the $d^*$ state has a mass of about $2.38-2.42$ GeV, a root-mean-square radius (RMS) of $0.76-0.88$ fm, and a CC fraction of $66\%-68\%$. The last may cause a rather narrow width to $d^*$ which, together with the quantum numbers and our calculated mass, is consistent with the newly observed resonance-like structure ($M\approx 2380$ MeV, $\Gamma\approx 70$ MeV) in double-pionic fusion reactions reported by WASA-at-COSY Collaboration.
\end{abstract}


\maketitle


The ABC effect has drawn physicists' great attention since its observation in 1961 in the $pd$ reaction \cite{ABC}.  
%
In recent years, much experimental progress in exploring the nature of the ABC effect has been made. 
In 2009, the CELSIUS/WASA
Collaboration measured the most basic double-pionic fusion
reaction $pn\to d\pi^0\pi^0$ with an incident proton energy of
1.03 GeV and 1.35 GeV \cite{CW2009}, and found significant enhancements in the $\pi\pi$ invariant 
mass spectrum at $\pi\pi$ invariant mass below 0.32 GeV$^2$ and also in the $d\pi$ invariant mass spectrum at $\Delta$ resonance region. To 
accommodate these data as well as the energy dependence of the total cross section at
$\sqrt{s}<2.5$ GeV, 
the conventional $t$-channel $\Delta\Delta$ intermediate state is found to be not sufficient, and a new structure, 
namely an $s$-channel resonance with mass of about 2.36 GeV and width of
about 80 MeV, is expected. In 2011, the 
WASA-at-COSY Collaboration further measured the $pn\to d\pi^0\pi^0$ reaction with the beam energies of
$1.0-1.4$ GeV which cover the transition region of the
conventional $t$-channel $\Delta\Delta$ process \cite{WASA2011}. They found that neither the $t$-channel $\Delta\Delta$ process nor
the Roper resonance process can explain the data, and an $s$-channel resonance with quantum numbers
of $I(J^P)=0(3^+)$, mass of about 2.37 GeV and width of about
70 MeV is indeed needed to describe the data. Recently, the WASA-at-COSY Collaboration measured the
polarized $\vec{n}p$ scattering through the quasi-free process
$\vec{d}p \to p_{\rm spectator} np$ \cite{WASA2014,WASA2014-2}. By incorporating the newly measured
$A_y$ data into the SAID analysis, they obtained a pole in the
$^3D_3$-$^3G_3$ waves at $(2380 \pm 10) + i (40\pm 5)$ MeV, which
again supports the existence of a resonance, called $d^*$, as mentioned
in Ref.~\cite{WASA2011}. Further evidence of this resonance has also been reported in the quasi-free $np\to np\pi^0\pi^0$ reaction \cite{WASA2014-3}. Since its mass is
above the threshold of  $\Delta N\pi$ channel, while its width is
much smaller than the decay width of $\Delta$, this
resonance must be a very interesting state involving new physical mechanisms and it is obviously worthwhile investigating.

Theoretically, the possibility of the existence of dibaryon states was first proposed in 1964 by Dyson and Xuong based on SU(6) symmetry \cite{Dyson64}. Since then, extensive efforts have been given in
exploring the possible existence of a $\Delta\Delta$ dibaryon on hadronic degrees of freedom. However, no convincing results have ever been released yet. Since the birth of quark
model, dramatic progresses on this aspect were pouring
in. In 1980, by analyzing the characteristics of the
one-gluon-exchange (OGE) interaction between quarks, Oka and Yazaki pointed out that in all non-strange
baryon-baryon (BB) systems, the $\Delta\Delta$ system with
$I(J^P)=0(3^+)$ is the only one in which the effective BB
interaction induced by OGE shows an attractive feature \cite{OY80}.
In the following years, by including the interaction between the
quark field and chiral field into the constituent quark model,
one successfully reproduced the data of the nucleon-nucleon (NN) interaction and the
binding energy of the deuteron \cite{ZYS97}, which would provide a
much reliable platform to predict the structures of dibaryons in the quark degrees of freedom. In 1999, we carefully performed a dynamical study of the $\Delta\Delta$ system in the quark degrees of freedom within the framework of Resonating Group Method (RGM), with the hidden color (CC) channel being properly taken into account \cite{YZYS99}. Where, by
employing the chiral SU(3) quark model with a set of
reasonable model parameters which can reproduce the NN
scattering phase shifts at relatively lower energies and the binding
energy of deuteron, we found that the quark-exchange effect in the
$\Delta\Delta$ system with  quantum number  $I(J^P)=0(3^+)$ is so important (see also Ref.~\cite{LSZY2001}) that the system should be bound in
nature with a binding energy of about $20-50$ MeV in a single
$\Delta\Delta$ channel calculation. The coupling to the CC channel was also intensively studied and found to play an important role in the binding behavior of the
system. It offers an additional binding energy of over 20 MeV
to the system, and thus the binding energy of $d^*$, relative to the
threshold of $\Delta\Delta$ channel, would run up to $40-80$ MeV. Unexpectedly,
our predicted mass and quantum numbers are quite close to the new observation
released recently by WASA-at-COSY Collaboration \cite{WASA2014}. 

In this work, we perform a further elaborate investigation of the $\Delta\Delta-$CC system within the framework of RGM in a chiral quark model. Besides the binding energy, we concentrate on a detailed study of the relative wave function of the $\Delta\Delta-$CC system, which is crucial to the understanding of the structure and decay property of $d^*$. We find for the first time based on a microscopic calculation that the  $d^*$ has a CC component of about 2/3, which indicates that $d^*$ is a hexaquark-dominated exotic state. The large CC configuration is quite helpful for us to understand that $d^*$, although locating above the thresholds of $\Delta N \pi$ and $NN\pi\pi$ channels, has a relatively narrow width as will be discussed later in more detail. In the most recent experimental papers by WASA-at-COSY Collaboration and SAID Data Analysis Center \cite{WASA2014-2,WASA2014-3}, our scenario for $d^*$ present here has been cited as a plausible explanation of the resonance observed in the double-pionic fusion reaction, as both our calculated binding energy and the expected width from our picture of $d^*$ are in agreement with the experimental findings.

The interaction between $i$-th quark and $j$-th quark in our chiral quark model reads
\begin{equation}
V_{ij} = V_{ij}^{\rm OGE} + V_{ij}^{\rm conf} + V_{ij}^{\rm ch},
\end{equation}
where $V_{ij}^{\rm OGE}$ is the OGE interaction which describes the short-range perturbative QCD behavior, and $V_{ij}^{\rm conf}$ the confinement potential describing the long-range non-perturbative QCD effects. $V_{ij}^{\rm ch}$ is the chiral field induced quark-quark interaction which provides the medium-range non-perturbative QCD effects, and to test the model dependence of our results, we employ two different models for this interaction. In the chiral SU(3) quark model, $V_{ij}^{\rm ch}$ reads
\begin{equation}
V_{ij}^{\rm ch} = \sum_{a=0}^8 \left(V_{ij}^{\sigma_a} + V_{ij}^{\pi_a} \right),
\end{equation}
and in the extended chiral SU(3) quark model, $V_{ij}^{\rm ch}$ reads
\begin{equation}
V_{ij}^{\rm ch} = \sum_{a=0}^8 \left(V_{ij}^{\sigma_a} + V_{ij}^{\pi_a} + V_{ij}^{\rho_a} \right),
\end{equation}
with $\sigma_a$, $\pi_a$ and $\rho_a$ ($a=0,1,\cdots,8$) being the scalar, pseudo-scalar and vector nonet fields, respectively. Note that the OGE will be largely reduced when vector-meson exchanges are included, i.e. the short-range interaction mechanisms are quite different in these two models. We refer readers to Refs.~\cite{ZYS97,DZYW2003} for further details.

\begin{table}[tb]
\caption{\label{tab:dbe} Binding energy, root-mean-square radius (RMS) of 6 quarks, and fraction of channel wave function for deuteron in chiral SU(3) quark model and extended chiral SU(3) quark model with ratio of tensor coupling to vector coupling f/g=0 and f/g=2/3 for vector meson fields. }
\begin{tabular*}{\columnwidth}{@{\extracolsep\fill}lrrr}
 \hline\hline
    &  SU(3) & \multicolumn{2}{c}{Ext. SU(3)}  \\ \cline{3-4}
    &            &  (f/g=0)       & (f/g=2/3)     \\ \hline
 Binding energy (MeV) &  2.09  &  2.24  &  2.20 \\ 
 RMS of $6q$ (fm) &    1.38   &  1.34  &   1.35 \\ 
 Fraction (NN)$_{L=0}$ ($\%$) & 93.68 &  94.66  & 94.71 \\
 Fraction (NN)$_{L=2}$ ($\%$) & 6.32  &  5.34  &  5.29 \\  \hline\hline
\end{tabular*}
\end{table}

The parameters of both models are fixed by fitting the energies of the octet and decuplet baryon ground states, the NN scattering phase shifts in the low energy region ($\sqrt{s} \leqslant 2m_N+200$ MeV) and the binding energy of deuteron \cite{ZYS97,DZYW2003}. See Table~\ref{tab:dbe} for the binding energy, root-mean-square radius (RMS) of $6q$, and the fraction of each partial wave for deuteron obtained from our models. They are all quite reasonable.

\begin{table*}[tb] 
\caption{\label{tab:dstarbe} Binding energy, root-mean-square radius (RMS) of 6 quarks, and fraction of channel wave function for $d^*$ in chiral SU(3) quark model and extended chiral SU(3) quark model with ratio of tensor coupling to vector coupling f/g=0 and f/g=2/3 for vector meson fields.}
\begin{tabular*}{\textwidth}{@{\extracolsep\fill}lrrrrrr} 
 \hline\hline
   & \multicolumn{3}{c}{$\Delta\Delta$ ($L=0,2$) } & \multicolumn{3}{c}{$\Delta\Delta\,-\,$CC  ($L=0,2$) } \\[1pt] \cline{2-4} \cline{5-7}
   & SU(3) & Ext. SU(3) & Ext. SU(3) & SU(3) & Ext. SU(3) & Ext. SU(3) \\
   &            &   (f/g=0)  &  (f/g=2/3)  &    &   (f/g=0)  &  (f/g=2/3)   \\ \hline
 Binding energy (MeV) &  28.96  &  62.28  &  47.90 & 47.27 & 83.95 & 70.25 \\ 
 RMS of $6q$ (fm) &   0.96   &  0.80  &  0.84  & 0.88 & 0.76 & 0.78 \\ 
 Fraction ($\Delta\Delta$)$_{L=0}$ ($\%$) & 97.18 &  98.01  & 97.71 &  33.11 & 31.22 & 32.51 \\
 Fraction ($\Delta\Delta$)$_{L=2}$ ($\%$) & 2.82  & 1.99  &  2.29 & 0.62 & 0.45 &  0.51 \\ 
 Fraction (CC)$_{L=0}$ ($\%$) &   &    &   &  66.25 & 68.33 & 66.98 \\
 Fraction (CC)$_{L=2}$ ($\%$) &   &    &   &  0.02 & 0.00 & 0.00 \\  
 \hline\hline
\end{tabular*}
\end{table*}

With all the parameters being properly fixed, we are eager to a investigation for the properties of the $\Delta\Delta$-CC system without introducing any new adjustable parameters. Here the hidden color channel CC with isospin $I=0$ and spin $S=3$ is built as
\begin{equation}
\Ket{CC}_{IS=03} \equiv - \frac{1}{2} \Ket{\Delta\Delta}_{IS=03} + \frac{\sqrt{5}}{2} {\cal A}^{\rm sfc} \Ket{\Delta\Delta}_{IS=03},
\end{equation}
with ${\cal A}^{\rm sfc}$ being the antisymmetrizer in spin-flavor-color space. We dynamically calculate the binding energy of this  system  by solving the RGM equation for a bound state problem, and then discuss the structure of this bound state via a systematic analysis of fractions of each channel in the resultant relative wave function. The results obtained in chiral SU(3) quark model and the extended chiral SU(3) quark model with the ratios of the tensor coupling to vector coupling f/g=0 and f/g=2/3 for vector meson fields are listed in Table~\ref{tab:dstarbe}, where the values for RMS of 6 quarks are also shown.

One sees from Table~\ref{tab:dstarbe} that the $\Delta\Delta$ state with $I(J^P)=0(3^+)$ has a binding energy of about $30-60$ MeV and a RMS of about $0.80-0.96$ fm in a $\Delta\Delta$ ($L=0,2$) double-channel calculation. The coupling to the CC channel will further result in an increment of about $20$ MeV to the binding energy and a considerable decrement of the RMS, and finally, the mass of this bound state will reach to $2.38-2.42$ GeV and the RMS will shrink to $0.76-0.88$ fm. This clearly shows that $d^*$ is a $\Delta\Delta$-CC deeply bound and compact state where the coupling to the CC channel plays a significant role. 

\begin{figure*}[tb]
\centering
\subfigure[~Deuteron]{
\label{fig:subfig:a} 
\includegraphics[width=0.48\textwidth]{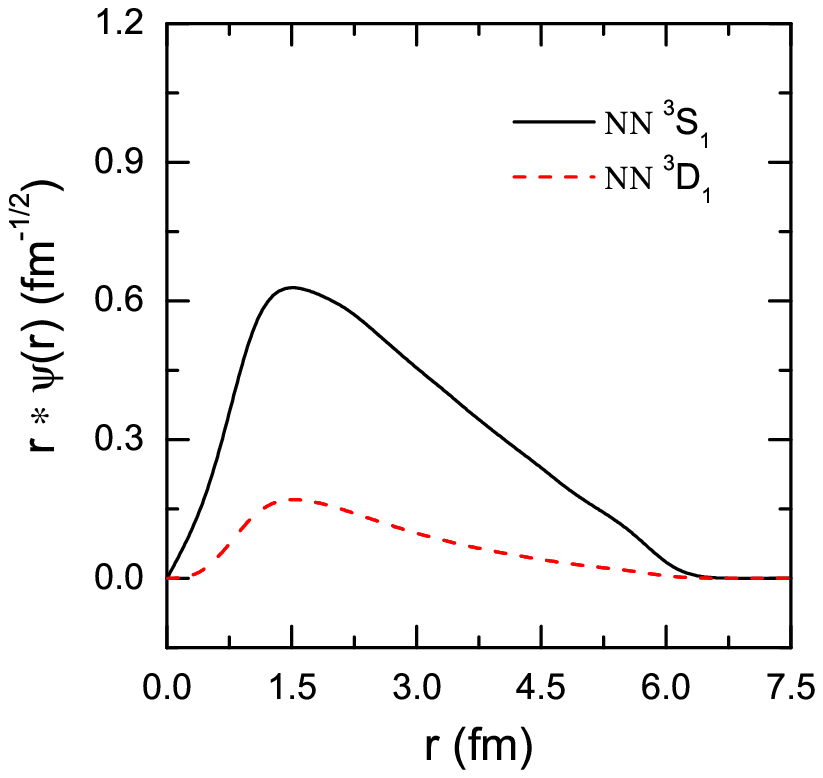}}
\subfigure[~$d^*$]{
\label{fig:subfig:b} 
\includegraphics[width=0.48\textwidth]{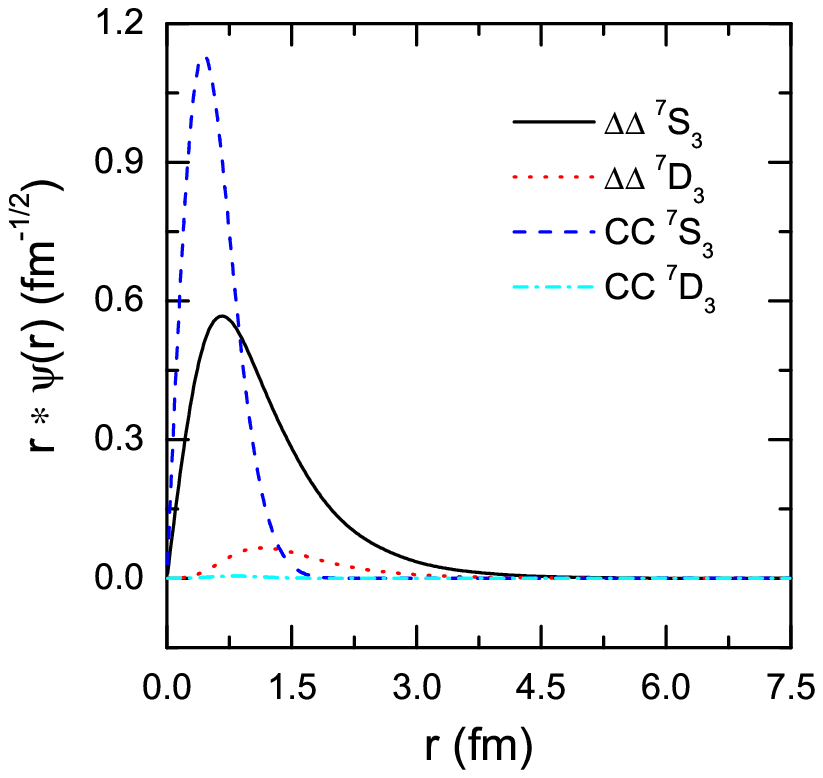}}
\caption{Relative wave functions in the extended chiral SU(3) quark model with f/g=0 for deuteron (left) and $d^*$ (right).}
\label{fig:wf} 
\end{figure*}

Apart from the RMS, more information about the ``size" of $d^*$ can be acquired from its spacial distribution feature as shown in Fig.~\ref{fig:wf}, where the partial-wave projected relative wave function in physical basis as a function of the distance between two physical states in the extended chiral SU(3) quark model with f/g=0 is plotted. The results in other two models are similar. For comparison, the results for deuteron are also plotted. Here the relative wave function in physical basis, namely the channel wave function in the quark cluster model, is defined in the usual way as \cite{Kusainov91,Glozman93,Stancu97}
\begin{equation}
\psi_{\rm BB}\!\left({\bm r}\right)  \equiv \Braket{ \phi_{\rm B}\!\left( {\bm \xi}_1, {\bm \xi}_2 \right) \phi_{\rm B}\!\left( {\bm \xi}_4, {\bm \xi}_5 \right) | \Psi_{6q} }, \label{eq:ch-wf-1}
\end{equation}
with B standing for $\Delta$ or C and $\Psi_{6q}$ being the six-quark wave function from RGM calculation,
\begin{equation}
\Psi_{6q} =\sum_{{\rm B}=\Delta,{\rm C}} {\cal A} \left[ \phi_{\rm B}\!\left( {\bm \xi}_1, {\bm \xi}_2 \right) \phi_{\rm B}\!\left( {\bm \xi}_4, {\bm \xi}_5 \right) {\eta_{{\rm B}{\rm B}}\!\left({\bm r}\right)} \right]. \label{eq:ch-wf-2}
\end{equation}
In Eqs.~(\ref{eq:ch-wf-1}) and (\ref{eq:ch-wf-2}), $\phi_{\rm B}$ is the antisymmetrized internal wave function for cluster B, with ${\bm \xi}_i$ ($i=1,2,4,5$) being the internal Jacobi coordinates for corresponding cluster; ${\cal A}$ is the antisymmetrizer for quarks from different clusters required by Pauli exclusion principle; and ${\eta_{{\rm B}{\rm B}}\!\left({\bm r}\right)}$ is the relative wave function between two clusters of B, which in present investigation is completely determined by the interacting dynamics of the whole six-quark system.
From Fig.~\ref{fig:wf} one  sees that $d^*$ is rather narrowly distributed and it has a maximal distribution located around $0.7$ fm for $\Delta\Delta$ ($L=0$) and $0.4$ fm for CC ($L=0$), respectively. While deuteron is widely distributed with a maximal distribution located around $1.4$ fm.

It should be mentioned that an even more interesting thing in this investigation is to study the fractions of relative wave functions for each individual channel, which will help us get a further understanding of the structure of $d^*$. Our extracted fractions of relative wave functions for $\Delta\Delta$ and CC channels are tabulated in Table~\ref{tab:dstarbe}. Of great interest is that one observes that the fraction of the CC channel in $d^*$ is about $66\%-68\%$. Note according to symmetry, a pure hexaquark state of $\Delta\Delta$-CC system with isospin $I=0$ and spin $S=3$ reads
\begin{equation}
[6]_{\rm orb} [33]_{IS=03} = \sqrt{\frac{1}{5}} \Ket{\Delta\Delta}_{IS=03} + \sqrt{\frac{4}{5}} \Ket{\rm CC}_{IS=03},
\end{equation}
which indicates that the fraction of CC channel in a pure hexaquark state is $80\%$. It is thus fair to say that $d^*$ is a hexaquark-dominated state as it has a CC configuration of $66\%-68\%$. This finding is of great interest. It helps us to understand that $d^*$, although locating above the thresholds of the $\Delta N \pi$ and $NN\pi\pi$ channels, has a relatively narrow width since the CC component cannot subject to a direct break-up decay. Actually, it can only decay to colorless hadrons via the re-combination processes of six color quarks together with quark-antiquark pair creation processes, which are highly suppressed and have much lower probability than those direct break-up decay processes. A conjecture that $d^*$ should have an unconventional origin and the CC configuration will suppress its decay width has also been proposed by Bashkanov, Brodsky and Clement in Ref.~\cite{BBC2013}. Our microscopic calculation presented here supports their argument and our dynamical results show that $d^*$ has about 2/3 hidden-color configurations. However, a detailed calculation of the decay width of $d^*$ needs to take into account both the kinematic effects and the effects from the dynamical structure of $d^*$. In the literature the former has been considered by using a simple momentum dependent prescription for $d^*$’s decay width \cite{HPW2014}. Now an investigation of the later also becomes possible, because the relative wave functions of $d^*$ is available in the present paper. Our work along this line is in progress.

It should particularly be stressed that the above-mentioned features of $d^*$ are due to both the quark exchange effect and the short-range interaction being attractive in the $\Delta\Delta$-CC system. As we have pointed out in Ref.~\cite{LSZY2001}, the quark exchange effect is highly dependent on the quantum numbers of the system. Opportunely, the $\Delta\Delta$ system with $I(J^P)=0(3^+)$ is one of the few systems which have strong quark exchange effect that drags two baryons together to form a compact state. As for the interaction property, Oka and Yazaki claimed in 1980 that in all the non-strange two-baryon systems, the $\Delta\Delta$ system with $I(J^P)=0(3^+)$ is the only one in which the OGE provides strong attraction in short range \cite{OY80}. In our chiral SU(3) quark model, the OGE indeed provides strong short-range attraction. In the extended chiral SU(3) quark model, although the OGE is largely reduced, the short-range interaction is still attractive and the attraction is even much stronger, as the vector meson exchanges (VMEs) are also strongly attractive in short range. Considering both mentioned facts, the $\Delta\Delta$ system with $I(J^P)=0(3^+)$ is certainly deeply bound and it would couple to the CC channel strongly as two interacting $\Delta$s are dragged closer enough by strong attraction. As a comparison, the NN $^3S_1$ partial wave has a rather different feature. In this partial wave, the quark exchange effect is very week (almost negligible), and moreover, the short-range interaction stemming from OGE and VMEs are all repulsive. Therefore, the NN $^3S_1$ partial wave can only get attraction in the medium-  and long-range through $\sigma$ and $\pi$ meson exchanges, and as a result, the deuteron is loosely bound and hardly couples to CC channel. We emphasize that the $\Delta\Delta$ system with $I(J^P)=0(3^+)$ is a highly special system where both the quark exchange effect and the short-range interaction are attractive, which makes this system deeply bound and promotes a strong coupling to the CC channel.

In summary, it is the first time one finds based on a microscopic calculation with no additional parameters besides those fixed already in the study of  NN scattering phase shifts that $d^*$ has about 2/3 hidden-color configurations and thus tends to be a hexaquark-dominated exotic state. It has a mass of about $2.38-2.42$ GeV, a root-mean-square radius of $0.76-0.88$ fm, and a CC fraction of $66\%-68\%$ which may cause a relatively narrow width of $d^*$. Our findings are consistent with the newly observed resonance-like structure ($M\approx 2380$ MeV, $\Gamma\approx 70$ MeV) in double-pionic fusion reactions reported by WASA-at-COSY Collaboration. We mention that if its character can be further verified, the $d^*$ will be the first hexaquark-dominated exotic state we have ever found, and it may open a door to new physical phenomena.

In literature there are several other interpretations of the recent WASA-at-COSY experiment.
Gal et al. carried out a $\pi N\Delta $ three-body-calculation \cite{Gal2013} and found the ${\mathcal
{D}_{03}}$ state as a dynamically generated pole at the right mass and slightly larger width (see Ref.~\cite{WASA2014-2} for a comment of the width). Huang et al. performed a coupled-channel quark model calculation and obtained an energy close to the observed value, but their calculated width is still too large \cite{HPW2014}.

The $\pi N \Delta$ three-body resonance scenario proposed by Gal {\it et al.} \cite{Gal2013} might have a relatively larger RMS than that from our picture. While measuring the ``size" of $d^*$ is rather involved, experiments in future might reach $d^*$ by electromagnetic transitions, which will give information about $d^*$ form factor and further tell us the real structure of the observed resonance \cite{C2014}.  We look forward to experimental progresses along this line.

We are grateful for constructive discussions with Prof. H. Clement
and Prof. R. L. Workman.


\begin{acknowledgments}
This work is partly supported by the National Natural Science Foundation of China under grants Nos. 11475181, 11105158, 11035006 and 11165005, the Key-project by the Chinese Academy of Sciences under project No. KJCX2-EW-N01. F.H. is grateful to the support of the One Hundred Person Project of the University of Chinese Academy of Sciences.
\end{acknowledgments}



\end{document}